\documentclass[aps, prl, reprint,
 amsmath,amssymb,
floatfix,
longbibliograpy
]{revtex4-2}

\usepackage{graphicx}
\usepackage{mathptmx}
\usepackage{dcolumn}
\usepackage[dvipsnames]{xcolor}
\usepackage[colorlinks=true, allcolors=NavyBlue]{hyperref}
\usepackage{bm}

\begin{document}

\preprint{FERMILAB-PUB-26-0461-AD}

\title{Experimental Demonstration of a Two-Dimensional Nonlinear
  Integrable System \\ in a Particle Accelerator}

\author{John Wieland}
\thanks{Corresponding author}
\email[e-mail: ]{jwieland@fnal.gov}
\altaffiliation[Previously at ]{Michigan State University, East Lansing, MI, USA}

\author{Alexander Valishev}
\author{Aleksandr Romanov}
\thanks{Corresponding author}
\email[e-mail: ]{aromanov@fnal.gov}
\author{Nikita Kuklev}

\author{Giulio Stancari}
\altaffiliation[Also at ]{Consortium for Advanced Science and
  Technology (CASE), University of Chicago, Chicago, IL, USA.}

\affiliation{Fermi National Accelerator Laboratory, Batavia, IL, USA}

\author{Sergei Nagaitsev}
\affiliation{Brookhaven National Laboratory, Upton, NY, USA}

\author{Sebastian Szustkowski}
\altaffiliation[Previously at ]{Northern Illinois University, Dekalb, IL, USA}

\affiliation{Los Alamos National Laboratory, Los Alamos, NM, USA}

\date{July 10, 2026}

\begin{abstract}
  A two-dimensional nonlinear integrable system was experimentally
  demonstrated at the Fermilab Integrable Optics Test Accelerator. 
  The system was implemented by inserting a special nonlinear
  magnet in a conventional accelerator lattice. We characterized the
  system by measuring lifetimes, transverse profiles and transverse
  oscillation frequencies of the 150-MeV electron beam as a function
  of the strength of the nonlinear insert. The measured shift of the
  working point and the amplitude-dependent detuning were consistent
  with theoretical predictions. We also observed the predicted
  bifurcation of the stable closed orbit. A striking consequence of
  the system's implementation was the possibility to operate the
  storage ring with integer tunes without lifetime degradation. This
  research opens up novel ways to design particle accelerators and to
  stabilize particle beams.
\end{abstract}

\maketitle

\section{Introduction}

A rich variety of phenomena in physics, engineering, chemistry,
ecology, economics and other fields can be modeled by the theory of
nonlinear dynamical systems~\cite{Lichtenberg:1992, Thompson:2002,
  Strogatz:2019, Kantz:2003}. These systems are characterized by
regular and chaotic motion in phase space and by sensitivity to
parameters and initial conditions. Of particular interest are
nonlinear integrable systems~\cite{Tabor:1989}. They possess as many
independent invariants of motion as the number of degrees of freedom
and exhibit regular behavior, while preserving the amplitude
dependence of oscillation frequencies typical of nonlinear
systems. However, they are in general difficult to describe
analytically and to realize experimentally.

Particle accelerators offer a compelling case to study the general
properties of dynamical systems and their practical applications. The
circulating particle beam travels in a vacuum chamber under the
influence of precisely controlled electromagnetic fields. Its
evolution takes place over relatively fast time scales and can be
passively monitored.  In addition, there is substantial freedom in the
design of the sequence of magnetic elements which define the confining
potential, usually referred to as the lattice.

As an application, nonlinear integrable optics (NIO) addresses the
basic problem of stable beam transport and focusing in particle
accelerators. The typical approach is based on the predictable and
stable dynamics of the harmonic oscillator. A sequence of dipole and
quadrupole magnets defines the linear potential, as described by
Courant and Snyder~\cite{Courant:1958}. In real machines, nonlinear
forces perturb the system. They can arise from magnet imperfections,
beam self fields and deliberate nonlinearities, introduced to
compensate for higher order effects. Most nonlinear dynamical systems
exhibit chaotic motion. In accelerators, this chaotic behavior
typically leads to phase-space degradation and beam loss. As beam
intensities, brightnesses and energies increase to satisfy
experimental demands, these beam losses pose a practical operational
limitation. Nonlinear integrable systems guarantee large regions of
bounded motion in phase space while generating an amplitude-dependent
detuning, which protects the beam from unwanted external perturbations
through the mechanism of Landau damping. Demonstrating their
feasibility opens up novel ways to design particle accelerators.

Nonlinear integrable optics for particle focusing has been considered
in the past. One-dimensional theoretical approaches date back to the
work of Orlov and McMillan~\cite{orlovAcceleratorsCollidingBeam1963,
  mcmillanThoughtsStabilityNonlinear1967}. Further work by Ruggiero
and Danilov~\cite{ruggieroIntegrabilityTwoDimensionalBeamBeam1982,
  danilovTwoExamplesIntegrable1997} focused on integrability using the
potential of the beam-beam interaction in a collider. The first
two-dimensional solution for transverse dynamics in an accelerator
using static electromagnetic potentials was proposed by Danilov and
Nagaitsev~\cite{danilovNonlinearAcceleratorLattices2010}. Their
approach can be summarized in three steps. First, the nonlinear
contribution to the Hamiltonian is separated from the linear
component. This has the effect of separating the accelerator lattice
into a nonlinear insert region and a linear matching lattice. Second,
the Hamiltonian is identified as a constant of motion by scaling the
nonlinear potential with the linear Courant-Snyder parameterization of
the motion. Finally, a direct search for a potential which generates a
second invariant of motion is undertaken. Assuming a second invariant
quadratic in the canonical momenta yields the Darboux
equation~\cite{darbouxProblemeMecanique1901}. By additionally
enforcing the condition that the potential satisfies Laplace's
equation, the potential can be realized with a static electromagnetic
field. Equation~\ref{eq:dnPot} shows the resulting Hamiltonian~$H$ of
the system, where $x$, $y$, $p_x$ and $p_y$ are the normalized
transverse coordinates and momenta and $t$ is a characteristic
geometric parameter of the nonlinear
potential~\cite{mitchellComplexRepresentationPotentials2019}:

\begin{eqnarray}
  \begin{split}
    H = & \frac{1}{2} (x^2 + y^2 + p_x^2 + p_y^2) + \\
    & t \cdot
    \operatorname{Re} \left[ \frac{x+iy}{\sqrt{1-(x+iy)^2}} \arcsin(x+iy) \right].
  \end{split}
  \label{eq:dnPot}
\end{eqnarray}

The main features of this NIO system are described below.  Their
experimental study is the object of this paper.

\section{Experimental Methods and Results}

\begin{figure}
  \centering
  \includegraphics[width=\linewidth]{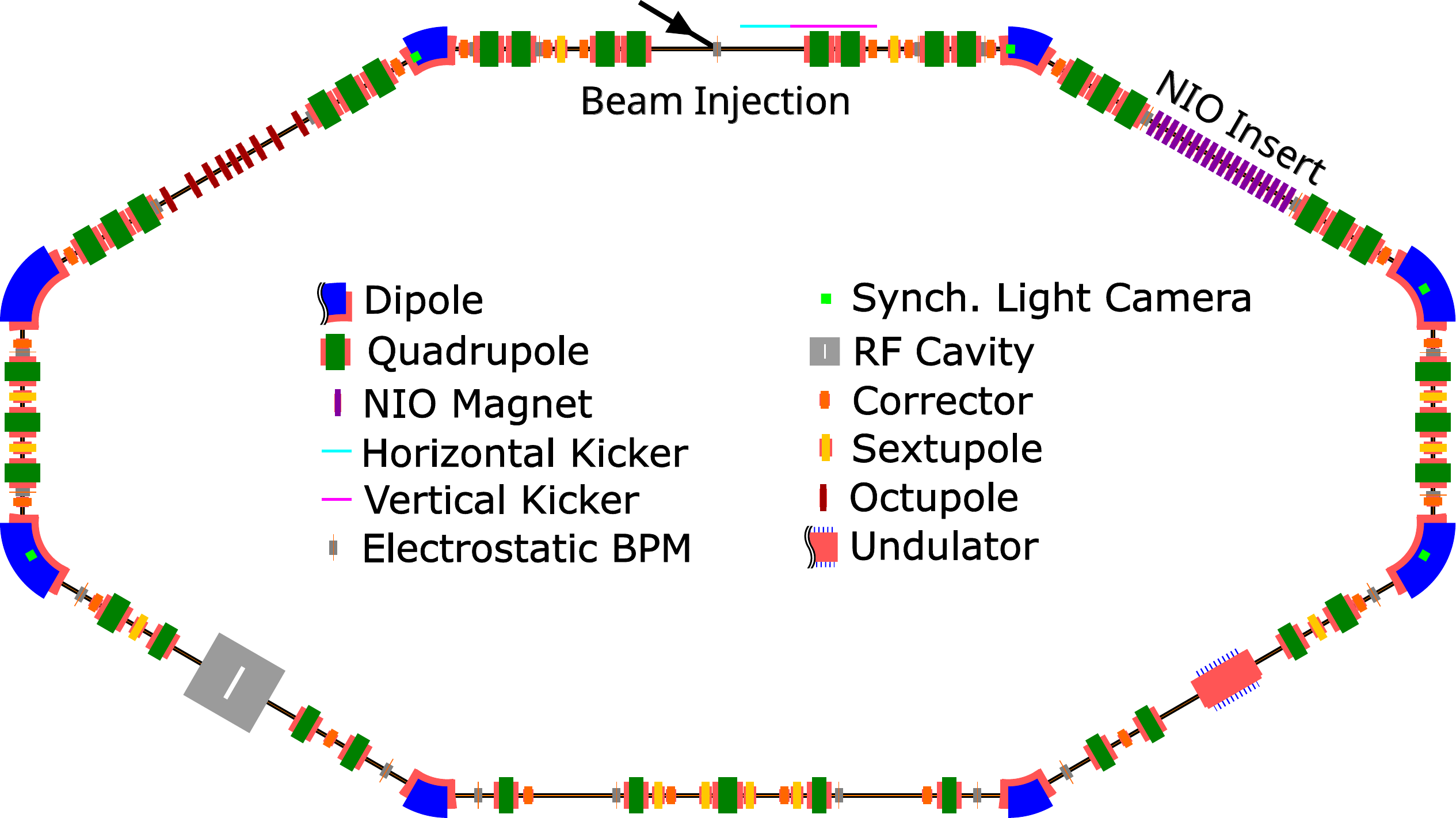}
  \caption{Schematic diagram of the IOTA storage ring.}
  \label{fig:iotaDiagram}
\end{figure}

\begin{figure}
  \centering
  \includegraphics[width=0.9\linewidth]{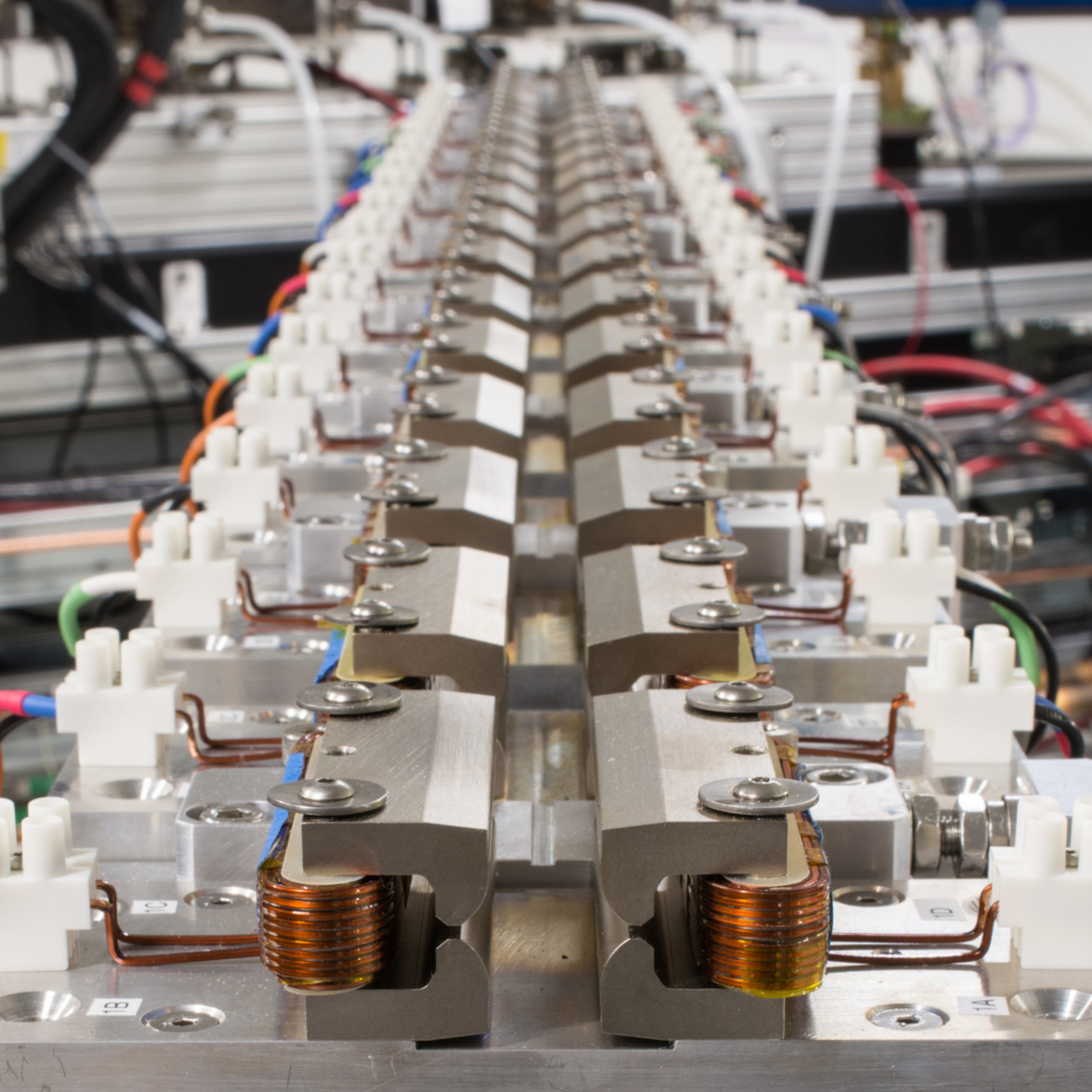}
  \caption{Photograph of the nonlinear magnetic insert.}
  \label{fig:nioPhoto}
\end{figure}

This system has been physically realized as a nonlinear insert in the
Fermilab Integrable Optics Test
Accelerator (IOTA)~\cite{antipovIOTAIntegrableOptics2017,
  Valishev:NIO-Proposal:2023}, sketched in
Figure~\ref{fig:iotaDiagram}. The linear lattice had a typical
strong-focusing design with conventional dipoles and quadrupoles. It
had enough flexibility to allow for granular tuning of the linear
Hamiltonian at the location of the NIO insert.  The ideal, continuous
NIO potential was piecewise approximated by a sequence of 18~custom
magnets (Figure~\ref{fig:nioPhoto}) placed in a 1.5-m-long section of
the ring. The poles of the magnets were shaped in order to generate
the proper potential according to their longitudinal position, leading
to 9~pairs of unique magnet poles arranged symmetrically about the
midpoint. The mechanical design and fabrication of the nonlinear
insert magnets and associated vacuum chamber are described in
Refs.~\cite{osheaMeasurementNonLinearInsert2013,
  mcnevinMechanicalDesignManufacturing2016}.

Experimentally, direct observation of the conservation of the
two predicted invariants was not the most sensitive and clean evidence
of the NIO system, for two main reasons. First, the residual ring
nonlinearities were amplitude dependent and limited the amplitude
range where the NIO system dominated the dynamics. Second, the
sensitivity of the turn-by-turn beam position monitors was
insufficient to resolve phase-space motion in the regions of
characteristic NIO behavior.

Nevertheless, three clear signatures of the NIO system were identified
and are presented here: (a) amplitude-dependent detuning of the
transverse oscillation frequencies; (b)~bifurcation of the stable
closed orbit; and (c)~stable operation of the ring at a working point
corresponding to the vertical integer resonance.

\subsection{Tune Shift and Amplitude-Dependent Detuning}

The tune of a harmonic system is the characteristic frequency of the
oscillations in phase space. Nonlinear systems are non-harmonic, and
the interpretation of the tune becomes more
nuanced~\cite{nagaitsevBetatronFrequencyPoincare2020}. For
experimental evaluation, the dominant frequency component of the
oscillations is interpreted as the tune. In a nonlinear system, there
is a characteristic amplitude dependence of the tune. Measurement of
the amplitude dependent detuning in a physical system can indicate the
form and magnitude of the nonlinearities.

For tune measurements in IOTA, two stripline
kickers~\cite{antipovStriplineKickerIntegrable2016}, one horizontal
and one vertical, were used to apply a one-turn deflection to the beam
centroid and excite coherent transverse oscillations with adjustable
amplitudes. Twenty-one beam position monitors around the ring sampled
the centroid position
turn-by-turn~\cite{eddyBeamInstrumentationFermilab2019}. The frequency
composition of this sample was then extracted to measure the
tune. Naturally, the beam was composed of many particles, each
oscillating at a different amplitude. Simulation studies indicated
that the measured tune of the centroid was an acceptable proxy for the
tune of a single particle at an identical amplitude, provided the
transverse emittance was small and sufficiently far from perturbing
resonances.

\begin{figure}
  \centering
  \includegraphics[width=\linewidth]{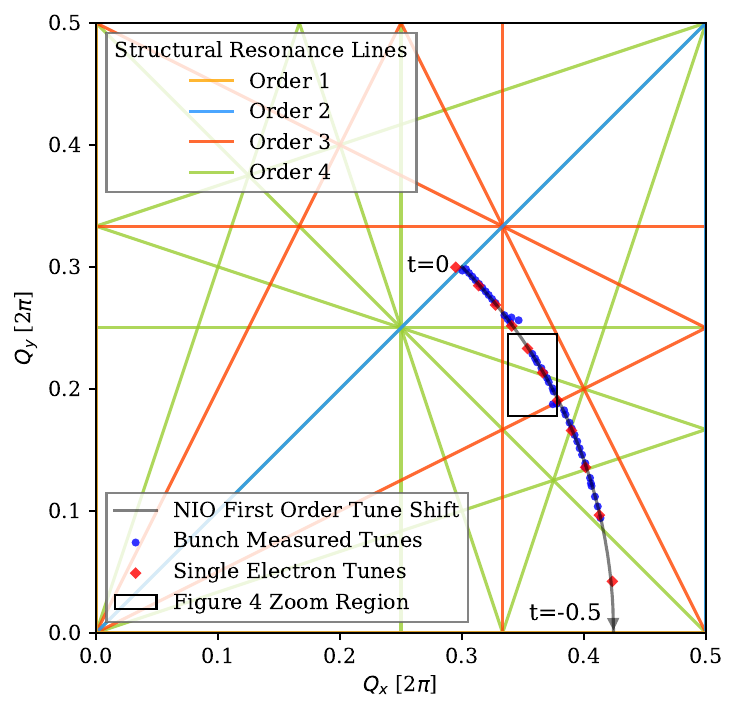}
  \caption{Measured and predicted working-point shift due to the
    first-order effect of the nonlinear insert. Uncertainties in tune
    measurements are smaller than the markers.}
  \label{fig:calibration}
\end{figure}

The impact of the NIO insert on the tune can be separated into two
effects: tune shift and amplitude-dependent detuning. The first-order
effect of the nonlinear potential is identical to that of a
quadrupole. The effect of this term is that of a shift of the central
working point, i.e.\ the tune of an equivalent linear system without
amplitude dependent detuning. The working point shift depends on the
strength of the dimensionless NIO scaling parameter~$t$, as given in
Equation~\ref{eq:qvt}~\cite{nagaitsevNonlinearOpticsPath2012}:

\begin{equation}
    Q_x, Q_y = Q_o \sqrt{1 \mp 2t}.
    \label{eq:qvt}
\end{equation}

Here $Q_o$ corresponds to the fractional component of the linear (or
``bare'') lattice working point of $Q_x = Q_y = 5.3$ with $t = 0$,
whereas $t = -0.5$ shifts the vertical working point to the integer
resonance at $Q_y = 5.0$. This shift is illustrated in tune space in
Figure~\ref{fig:calibration} as the gray arc. Measurements of the
working point were taken for different $t$-parameters by kicking the
beam. To minimize amplitude dependent effects, the smallest amplitudes
for which tune was resolvable were used. Quantitative fits of the
working point shift show good agreement with the model, and were used
to calibrate the actual $t$-parameter corresponding to a given current
setting of the nonlinear magnet.

In IOTA, it is also possible to circulate and monitor single
electrons. This remarkable regime enables high-precision measurements
of machine properties~\cite{Lobach:JINST:2022,
  romanovExperimental3dimensionalTracking2021}. Single electrons are
continuously kicked to small amplitudes by discrete emissions of
synchrotron radiation and by collisions with the residual
gas. Therefore, external excitations are not needed for tune
measurements. As a result, the working points from single-electron
measurements extend closer to the integer
resonance~\cite{romanovAnalysisEllipticIntegrable2025}, where the
amplitude-dependent detuning limits kicked-bunch experiments, as seen
in Figure~\ref{fig:calibration}.

\begin{figure}
  \includegraphics[width=\linewidth]{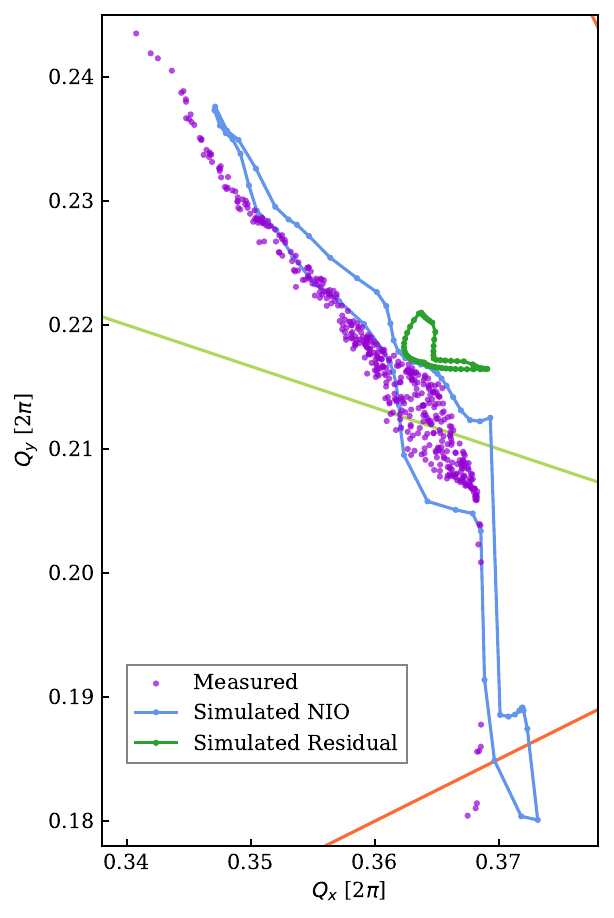}
  \caption{Comparison of tune footprints for experimental data and
    numerical simulations. The data points are in purple. Simulation
    footprints are represented by a boundary corresponding to the
    amplitudes of experimental kicks. The blue trace represents
    simulations with the NIO insert and the expected perturbative
    nonlinearities in the matching lattice. The green trace consists
    of simulations of just the perturbative nonlinearities and the 
    linear component of the NIO insert.}
\label{fig:detuning}
\end{figure}

The second effect of the NIO insert on the tune is the aforementioned
amplitude-dependent detuning. This effect produces a tune footprint
from the tune shift of all possible amplitudes, limited by the
the physical aperture
constrained by vacuum chamber size and by the dynamic aperture
defined by the maximum stable amplitudes of a nonlinear system.

The footprint of a system depends on all nonlinearities. In addition
to the NIO insert, the matching lattice of IOTA contains numerous
nonlinearities, some intentional and some residual. The intentional
nonlinearities stem from sextupole elements used to compensate for
energy-dependent focusing (chromaticity). The NIO Hamiltonian is
derived in two dimensions and chromaticity is a small unavoidable
effect of the three-dimensional structure of accelerator
dynamics. These perturbative sextupole terms drive corresponding
third-order resonances, reduce the dynamic aperture and dictate viable
working points. Residual nonlinearities in IOTA stem primarily from
kinematic effects in the main bending dipoles with short radii of
curvature and from the fringe fields at the edges of the quadrupole
magnets. These perturbative nonlinearities mainly affect the shape of
the footprint.

For the greatest footprint without substantial perturbation by the
sextupole terms, a working point corresponding to $t = -0.238$ was
selected. After measuring the tune footprint for the full available
range of amplitudes, comparative numerical simulations with the
residual nonlinearities similar to those in IOTA were
performed. Figure~\ref{fig:detuning} shows the measured and simulated
footprints. The systematic offset is explained by an imperfect
calibration of the experimental lattice, consistent with short-term
operational drifts. The nonlinear NIO detuning was verified to be the
dominant effect by comparing with simulations containing only the
linear component of the NIO insert. In this case, the detuning was
much smaller and its direction was opposite to the experimentally
measured values. The two tune-dependent effects suggest that we have a
strongly nonlinear system consistent with a dominant, well calibrated
NIO insert and perturbative nonlinearities in the matching lattice.

\subsection{Beam Lifetimes and Synchrotron Radiation Profiles:\\
  Stability at the Integer Resonance}

\begin{figure*}
  \includegraphics[width=0.8\textwidth]{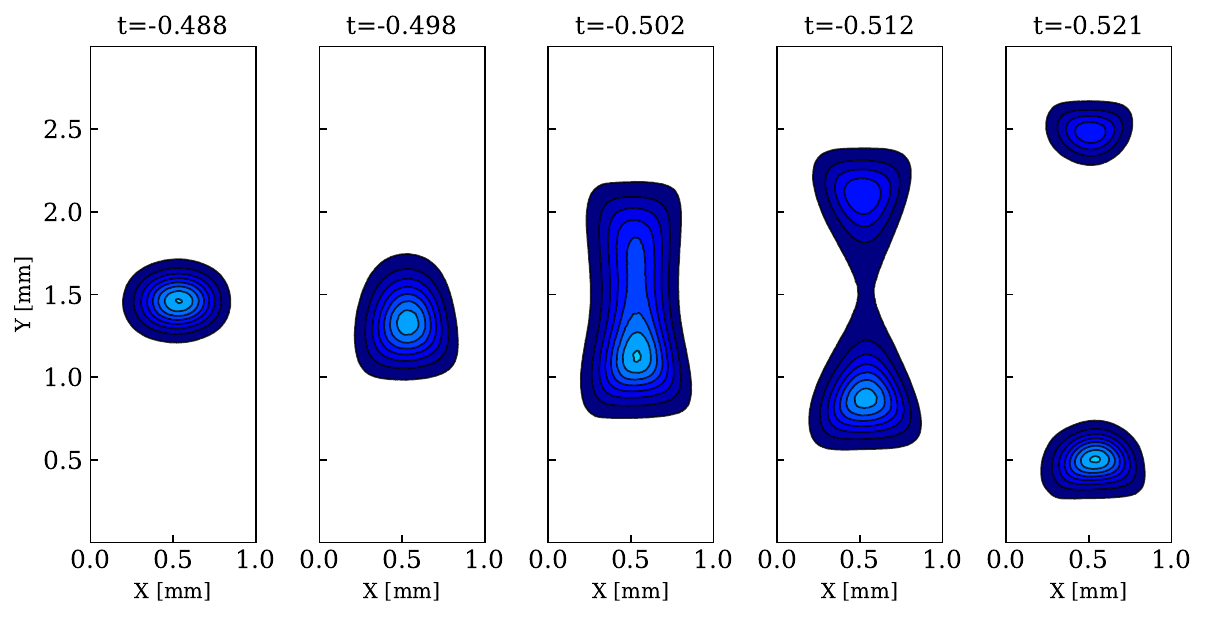}
  \caption{Measured transverse beam distributions while crossing the
    integer resonance.}
  \label{fig:integer}
\end{figure*}

\begin{figure}
  \centering
  \includegraphics[width=\linewidth]{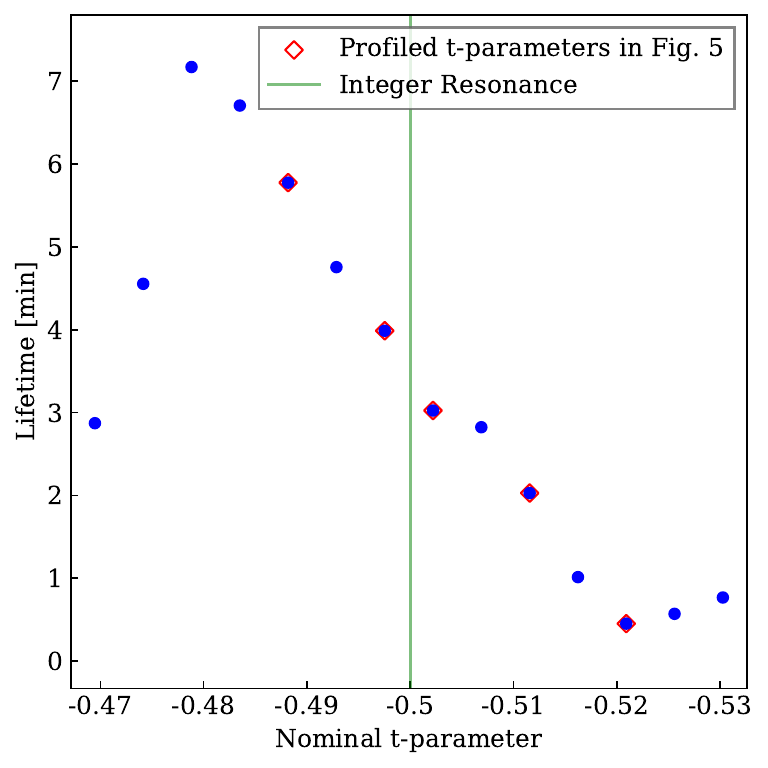}
  \caption{Measured beam lifetime vs.\ $t$-parameter strength. Uncertainties 
    in lifetime are smaller than the markers Points outlined in red squares 
    correspond to the beam distributions in Figure~\ref{fig:integer}.}
  \label{fig:lifetime}
\end{figure}

For a linear system, a transverse fractional tune of zero corresponds
to the integer resonance, a fundamentally unstable working point, due
to the effects of imperfect alignment of the closed orbit and of
perturbative nonlinearities \cite{conteIntroductionPhysicsParticle2008}.
For the NIO system, if $t = -0.5$, the vertical tune reaches the integer
resonance, as shown in Eq.~\ref{eq:qvt}. For NIO, this condition 
corresponds to a bifurcation
point instead of an instability. The stable fixed point on the axis of
the closed orbit becomes an unstable point, and two new stable fixed
points develop on the vertical axis and move symmetrically away from
the origin~\cite{mitchellBifurcationAnalysisNonlinear2020}. By
monitoring the beam lifetime at a working point on the integer
resonance, we evaluated the stability of the NIO system where a linear
system cannot operate.

To measure the transverse distribution of the beam, synchrotron
radiation in the bending dipoles was used. Radiation emission is
proportional to the beam density. It was sampled in five camera
stations around IOTA~\cite{eddyBeamInstrumentationFermilab2019}. The
radiation cone and dipole bending radii were small, so the imaged
radiation was dominated by the transverse distribution of the beam at
a well defined position within the bending magnet. For measurements at
the integer resonance, circulating beam was established and the
$t$-parameter was incremented in small steps.

As the integer resonance was approached, the perturbative
nonlinearities in the matching section of the ring dramatically
reduced the dynamic aperture, restricting the circulating beam to
small-amplitude particles. These amplitude limitations prevented
kicked-bunch tune measurements. Additionally, the derivative of the
tune with respect to the $t$-parameter grows near the integer
resonance condition, so systematic uncertainties in the calibration
become more pronounced. The combination of these effects prevented us
from identifying the integer resonance setpoint directly from tune
calibrations.

Instead, the change in transverse distribution was used to accurately
identify the integer resonance setpoint. By observing when the two
fixed points beyond $t = -0.5$ appeared, we could verify the location
of the integer resonance within the uncertainty of the step size in
$t$-parameter. Figure \ref{fig:integer} shows the measured bifurcation 
of the original closed orbit into two new closed orbits about the 
stable fixed points. Due to the slight asymmetry of the circulating 
beam with respect to the new fixed points, we see that more beam is 
captured into the lower beamlet. This indicates that the two closed 
orbits are independent, consistent with theoretical predictions for the 
IOTA design lattice. The lifetime and symmetry of the distribution were 
optimized by adjusting the linear focusing (phase advance, beam waist, 
and closed orbit), to compensate for imperfect calibration of the bare 
lattice and for short-term drifts.

At each setpoint, the beam intensity was measured over 10~s to
evaluate the beam lifetime, presented in Figure~\ref{fig:lifetime}. At
the setpoints which lie on either side of the integer resonance, the
beam lifetime was about 3~minutes, which corresponds to
1.3$\times10^{9}$ turns. This indicates asymptotic stability conferred
by the NIO optics for particles at the integer resonance, a
fundamentally unstable condition in the underlying linear system. The
general trend in lifetime, caused by perturbative nonlinearities, was
unaffected by the integer resonance. Beyond the integer resonance, we
observed the expected formation of two independent, stable
beamlets. These are unique characteristics of the NIO system.

\section{Discussion and Conclusions}

A two-dimensional nonlinear integrable system was implemented in an
accelerator lattice. The detuning of the system and the evolution of
stable fixed points were consistent with theoretical predictions. The
beam was observed at the linear integer resonance where stability is derived
entirely from the nonlinear insert.  The system was robust against
perturbations of the linear lattice, both unintentional (from
imperfect lattice calibration and short-term drifts, for instance) and
deliberate, such as changes in phase advance. The most significant
impact on the NIO system stemmed from sextupole
nonlinearities. Further efforts will focus on understanding the
interaction of perturbative nonlinearities on the stability of the
system in practical configurations. The research program will continue
with the study of nonlinear integrable systems and
space-charge-dominated beams, based on the capability of IOTA to store
2.5-MeV protons.

These studies open up novel approaches to accelerator design and beam
control, with implications for the operation of current high-intensity
accelerators and for the design of future facilities for the study of
rare processes in nuclear and particle physics.

\begin{acknowledgments}
  The authors would like to thank the Fermilab colleagues who made
  these experiments possible, in particular D.~Broemmelsiek,
  K.~Carlson, N.~Eddy, D.~Edstrom, J.~Jarvis, D.~MacLean, J.~Ruan, J.~Santucci,
  and T.~Thompson. The authors would also like to acknowledge the support of 
  P.~Ostroumov and the Michigan State University Accelerator Science and
  Engineering Traineeship. This manuscript has been authored by Fermi Forward
  Discovery Group, LLC under Contract No.~89243024\-CSC\-000002 with
  the U.S.\ Department of Energy, Office of Science, Office of High
  Energy Physics. Work supported in part by the DOE General
  Accelerator Research and Development (GARD) Program, the
  University of Chicago Consortium for Advanced Science and
  Engineering (CASE) and by Brookhaven Science Associates, LLC under 
  Contract No.~DE-SC0012704 with the U.S.\ Department of Energy.
\end{acknowledgments}

\bibliography{2026_NIO_PRL}

\end{document}